\documentclass{article}
\usepackage[a4paper, total={6in, 8in}]{geometry}
\usepackage{graphicx} 
\usepackage{amsmath}
\usepackage{amsfonts}
\usepackage[labelfont=bf]{caption}
\usepackage{subcaption}

\usepackage{biblatex}
\usepackage[table]{xcolor} 

\usepackage{array}
\newcommand{\PreserveBackslash}[1]{\let\temp=\\#1\let\\=\temp}
\newcolumntype{C}[1]{>{\PreserveBackslash\centering}p{#1}}
\newcolumntype{R}[1]{>{\PreserveBackslash\raggedleft}p{#1}}
\newcolumntype{L}[1]{>{\PreserveBackslash\raggedright}p{#1}}

\usepackage{multirow}

\definecolor{LightGray}{gray}{0.9}
\definecolor{seabornBlue}{RGB}{76,114,176}
\definecolor{seabornGreen}{RGB}{85,168,104}
\definecolor{seabornRed}{RGB}{196,78,82}
\definecolor{seabornOrange}{RGB}{226, 74, 51}

\usepackage[colorlinks=true, linkcolor=blue, urlcolor=blue, citecolor=black]{hyperref}

\usepackage{booktabs}
\usepackage{orcidlink}

\usepackage{tikz}
\usetikzlibrary{tikzmark}
\usetikzlibrary{calc}

\renewcommand*{\cite}{\parencite} 

\usepackage{lineno}

\begin{document}

\begin{flushleft}
{\Large
\textbf\newline{Generative AI-based data augmentation for improved bioacoustic classification in noisy environments}
}
\newline
\\
Anthony Gibbons 
\orcidlink{0000-0001-8479-4669}
\textsuperscript{1*},
Emma King
\orcidlink{0000-0002-8755-3884}
\textsuperscript{2},
Ian Donohue
\orcidlink{0000-0002-4698-6448}
\textsuperscript{2},
Andrew Parnell
\orcidlink{0000-0001-7956-7939}
\textsuperscript{3}
\\
\bigskip
\textbf{1} Hamilton Institute, Department of Mathematics and Statistics, Maynooth University, Kildare, Ireland
\\
\textbf{2} Zoology,
School of Natural Sciences, Trinity College Dublin, Dublin, Ireland
\\
\textbf{3} School of Mathematics and Statistics, University College Dublin, Dublin, Ireland

\bigskip

* anthony.gibbons.2022@mumail.ie

\end{flushleft}

\section*{Abstract}
    Obtaining data to train robust artificial intelligence (AI)-based models for species classification can be challenging, particularly for rare species. Data augmentation can boost classification accuracy by increasing the diversity of training data and is cheaper to obtain than expert-labelled data. However, many classic image-based augmentation techniques are not suitable for audio spectrograms.
    We investigate two generative AI models as data augmentation tools to synthesise spectrograms and supplement audio data: Auxiliary Classifier Generative Adversarial Networks (ACGAN) and Denoising Diffusion Probabilistic Models (DDPMs). The latter performed particularly well in terms of both realism of generated spectrograms and accuracy in a resulting classification task. 
    Alongside these new approaches, we present a new audio data set of 640 hours of bird calls from wind farm sites in Ireland, approximately 800 samples of which have been labelled by experts. Wind farm data are particularly challenging for classification models given the background wind and turbine noise.
    Training an ensemble of classification models on real and synthetic data combined compared well with highly confident BirdNET predictions. Each classifier we used was improved by including synthetic data, and classification metrics generally improved in line with the amount of synthetic data added.
    Our approach can be used to augment acoustic signals for more species and other land-use types, and has the potential to bring about advances in our capacity to develop reliable AI-based detection of rare species. Our code is available at \url{https://github.com/gibbona1/SpectrogramGenAI}.

\textbf{Keywords:} Bioacoustics, Data Augmentation, Deep Learning, Generative Models, Generative Adversarial Networks, Passive Acoustic Monitoring, Species Recognition, Stable Diffusion

\section{Introduction}\label{ch:intro}

Species monitoring is a major component of assessing biodiversity and ecosystem health, serving as a barometer of environmental change and conservation needs. The dynamics and diversity of bird communities are key indicators of ecosystem health and environmental change \cite{https://doi.org/10.1111/j.1365-2664.2011.02094.x}. Birds provide valuable ecosystem services, which include limiting pest populations \cite{GARCIA2024108927}, seed dispersion \cite{https://doi.org/10.1002/ecs2.2673} and pollination \cite{Toon2014}. The fields of bioacoustics and ecoacoustics have long studied the behaviour and diversity of bird species activity via sound. Acoustic monitoring has become increasingly popular due to increased efficiency, and reduced cost, compared to traditional methods \cite{https://doi.org/10.1111/2041-210X.12060, https://doi.org/10.1111/gcb.17067}. However, many concerns remain surrounding the accuracy and ability of software to identify bird species from acoustic recordings \cite{https://doi.org/10.1111/2041-210X.13101, ROSS2021107114}.

Computer vision developments have carried over to the field of bioacoustics to detect species via sound. Audio is represented as spectrograms in order to use image classification methods \cite{Stowell2022-vh}. Labelled data is required to train such classifiers, necessitating audio clips annotated to the degree of detail desired with, for example, content such as species \cite{Stowell_2014}, call type \cite{Bravo_Sanchez2021-ts}, individual animals \cite{Linhart2022}, or other audio events present. However, manual labelling of these datasets is both costly and time-consuming to obtain, requiring expert knowledge of bird vocalisations and skills in annotation software. Machine learning, in particular supervised learning, can automate large parts of such classification workflows, and requires high quality training data. Bioacoustics is still a nascent field and lacks standardised labelled datasets \cite{Stowell2022-vh} to facilitate training these models. To expand on the limited data available, data augmentation is often applied. However, not all data augmentation methods designed for images apply readily to spectrograms due to their inherent time-frequency structure \cite{Zhou2022}.

We investigate both Auxiliary Classifier Generative Adversarial Networks (ACGAN) \cite{odena2017acgan} and Stable Diffusion models  \cite{rombach2022highresolutionimagesynthesislatent} as data augmentation methods for enhancing bird call spectrogram classification models, with the aim of reducing the amount of domain-specific labelled data required to train species classifiers. Our key contributions include:
\begin{itemize}
    \item Application of Stable Diffusion to bird species monitoring. The model generates high-quality and diverse spectrograms of bird sound, as supported by both image and audio quality evaluation metrics. As far as we are aware, this is the first application of stable diffusion to bioacoustic classification of bird species.
    \item Improved classification. Each classifier we used found some improvement by including synthetic data. Classification metrics generally improved in line with the amount of synthetic data added.
    \item Ecoacoustic pipelines. We suggest generative models to enhance bioacoustic classification pipelines. Our Stable Diffusion model produces high-quality diverse spectrograms, improving bird song classification accuracy and reducing the need for expensive expert-labelled data.
\end{itemize}

Our paper is structured as follows. First, we give background on acoustic species classification and data augmentation and place our techniques in the context of existing data augmentation methods. After describing the details of our acoustic dataset, we then detail the audio pre-processing and the generative models we employed. We use the terms stable diffusion, diffusion models, Denoising Diffusion Probabilistic Models (DDPMs), and latent diffusion interchangeably once the stable diffusion model is introduced. We then provide a detailed description of our experimental design and results, and follow this up with in-depth discussion of our results, limitations and paths for future work.

\section{Existing Methods}\label{ch:background}

Passive Acoustic Monitoring (PAM) is a non-invasive technique where sensors are deployed in the field to record audio, usually on a pre-defined schedule, for weeks, months or years at a time without intervention. The ever-decreasing cost of recording equipment and data storage has allowed much larger PAM datasets to be collected \cite{birdclef-2024, DCASE2017challenge, https://doi.org/10.1007/s11284-017-1509-5}. The advancement of deep learning methods has continued to grow, allowing automatic species detection methods to achieve higher classification performance for the growing volume of data emerging via PAM. The scale and sophistication of these classifiers has advanced rapidly from presence/absence \cite{MacAodha2018} to classifying single digit \cite{doi:10.1080/09524622.2024.2354468} to tens and hundreds \cite{AKBAL2022101529, ZHONG2020107375} of species. BirdNET \cite{birdnet} and Perch \cite{perch}, both considered among the current state of the art in terms of bird classification capabilities, were trained on over 6,000 and 10,000 bird species globally, respectively.

The typical bird species classifier's input format will be as spectrograms, a visual representation of sound (see Figure~\ref{tab:real_synth}). The spectrograms are coloured by sound energy levels across time and frequency. These may be standard (linear-frequency), mel (to better align with human sound perception), or log-frequency spectrograms. Other audio representations used as input for classification include Scalogram \cite{app14010090}, MFCC \cite{9579881}, Gammatone Filterbank \cite{9005226}, Constant-Q Transform \cite{9023236} and the raw waveform itself \cite{lee2017rawwaveformbasedaudioclassification}.

While generalised solutions for bird species classification such as BirdNET and Perch are widely applicable, fine-tuned approaches are often desired for specific tasks \cite{https://doi.org/10.1111/2041-210X.14003}. The benefits include optimising hyperparameters, improving performance and robustness or reducing the number of parameters, which, in turn, reduces the size and computational cost of deploying such models \cite{jeddi2020simplefinetuningneedrobust,9645649}. Training custom classifiers is commonly done through techniques such as transfer learning \cite{ZHAO2024122807} and knowledge distillation \cite{Gou2021,hinton2015distilling}. These training processes necessarily require some amount of labelled training data from the specified task of interest. This can, in the case of acoustic data containing bird species, be time- and labour-intensive to obtain. When the amount of reliably labelled data is limited, additional strategies become appropriate to increase the utility of the available recordings.

Semi-supervised learning is a strategy to remedy class imbalance or issues with low volumes of high quality data, where labelled datasets are supplemented with unlabelled data \cite{NIPS2004_96f2b50b}. It is common, for example, to train a classifier on a small amount of labelled (human-verified) data, run through a large amount of unlabelled data, and take the highly confident predictions to be \emph{pseudo-labels}. Unsupervised learning models are also commonly used to generate these pseudo labels. The model is then retrained with both the labelled and pseudo-labelled data and, now that it has more examples to learn from, this often results in improved performance. Semi-supervised learning has been applied to bird call classification with few shot learning \cite{moummad2024selfsupervisedlearningfewshotbird}, beluga whale detection \cite{10.1121/10.0000921}, and individual bird recognition via object detection \cite{https://doi.org/10.1111/2041-210X.13436}. In our setting, the entirety of our training and validation datasets are \emph{unlabelled}, i.e. pseudo-labelled by highly confident (over 50\% confidence level) BirdNET predictions but not human verified. The test set contains highly confident human annotated data from the same sites, and our aim is for a robust model that performs well on this unseen test dataset.

Data augmentation is widely used to improve the performance of image classification models by artificially increasing the diversity of training data. Common augmentation techniques such as flipping, rotation, cropping, translation, and scaling/zooming help improve generalisation by providing varied perspectives of the same image content \cite{Shorten2019}. However, these methods are less applicable to spectrograms due to the inherent time ($x$) and frequency ($y$) structure of the latter \cite{gupta2021}. For example, excessive translation in the form of time or pitch shifting can distort the temporal correlation or pitch accuracy, leading to misleading representations of the audio content \cite{Zhou2022}. While techniques such as mixup \cite{zhang2018} and adding background noise \cite{Mu2021} are somewhat common for augmenting spectrograms, they also present challenges. Mixup, which interpolates between two samples, has shown promise \cite{2403.09598}, but can introduce temporal correlation artifacts or fail to maintain the original signal’s integrity. Additionally, adding background noise can degrade the quality of the signal, especially if a large volume of clean samples with birds present is not available, potentially hindering the model's ability to distinguish between the target sound and the noise \cite{kim2021specmixmixedsample}.

Generative models have emerged as a popular technology capable of creating diverse content, be it text, images, videos, or music/audio. There has been a recent surge in interest in generative AI with the release of Dall-E \cite{ramesh2021zeroshottexttoimagegeneration}, MidJourney \cite{midjourney} and Stable Diffusion \cite{rombach2022highresolutionimagesynthesislatent}, which are being used in personal and commercial projects of all scales. Synthetic image generation started with autoencoders \cite{JMLR:v11:vincent10a, kingma2022autoencodingvariationalbayes} and autoregressive models \cite{uria2016neuralautoregressivedistributionestimation}, before Generative Adversarial Networks (GANs) became the state of the art \cite{goodfellow2014gan}. Many more specific architectures based around GANs were designed and trained for their own specific use cases in image generation, such as GauGAN for generating images based on an initial semantic layout \cite{park2019semanticimagesynthesisspatiallyadaptive}. WaveGAN and SpecGAN generate audio in the form of raw waveforms and spectrograms respectively \cite{donahue2019adversarialaudiosynthesis}. A similar approach using mel spectrograms is melGAN \cite{kumar2019melgangenerativeadversarialnetworks}. Diffusion models are a more recent advancement that led to Stable Diffusion, involving both an autoencoder and a diffusion model on the latent representation of images. Latent diffusion models have begun to be used for spectrogram generation \cite{Howard2024fastai}. 

Our implementation of generative AI focuses on supplementing training data to improve classification performance that has been demonstrated in bird species classification \cite{https://doi.org/10.1111/2041-210X.14239}. In contrast, many generative AI uses are purely focused on image/data synthesis for direct use. ACGAN has specifically been applied for acoustic classification of bird species \cite{10.1121/10.0005202,dracgan} but we include here since it is a more widely used architecture and is faster at generating samples than diffusion models. Diffusion models have been used to improve classification performance of marine mammals \cite{10754260} but, as far as we are aware, this is the first implementation of them as augmentation methods for bird species classification, and the first use of either method for species classification at wind farms.

\section{Materials}\label{ch:materials}

In this section, we discuss the audio data collection setup and the steps required to generate pseudo-labelled training and validation sets using a state-of-the-art classification model. We use a set of human-labelled data using NEAL \cite{neal}, an audio annotation Shiny app, as an unseen test set to assess our classifiers.

For our study, audio data were collected from five wind farm sites across Ireland using the Song Meter Mini Bat recorder from Wildlife Acoustics \cite{smminibat}. The data were collected as part of the Nature+Energy Project, a research collaboration between academic
scientists, industry partners, and governmental agencies that focuses on
facilitating the enhancement of biodiversity on onshore windfarms in
Ireland \cite{GORMAN2024121814}. Sites were selected as representative of onshore windfarms across Ireland, and the habitats and land-uses in which they are typically co-located \cite{KING2025126618}.

The acoustic settings were configured as follows: a sample rate of 24 kHz was used, with a recording schedule of 5 minutes followed by a 15-minute non-recording interval to extend deployment time. The right channel gain (the left channel is the ultrasonic microphone) was set to 24 dB. The files were saved in~\verb|.wav| format. For this study, recordings from one recording cycle of each of the five wind farm sites were used, totalling approximately 640 hours of acoustic data. Deployments varied in length, as did the frequency of collecting batches of data stored on the SD cards. Some summary statistics about our data set are shown in Table~\ref{tab:data_info}.

\begin{table}[!ht]
    \caption{\textbf{A description of our wind farm audio dataset.}}
\centering
    \begin{tabular}{L{3cm}L{2cm}L{4.1cm}R{1.5cm}R{1.5cm}}
        \toprule
            Site & Location & Date range & \# files &  \# hours\\
        \midrule
        Carnsore & Wexford & 06 Jul 2022 - 23 Aug 2022 & 725 &   60 \\
        Cloosh Valley & Galway & 22 Dec 2022 - 19 Jan 2023 & 920 &   77 \\
        Rahora & Kilkenny & 07 Jan 2023 - 16 Apr 2023 & 2612 &  218 \\
        Richfield & Wexford & 05 Jul 2023 - 27 Jul 2023 & 1504 &  125 \\
        Teevurcher & Meath & 30 May 2022 - 26 Jun 2022 & 1921 &  160 \\
        \midrule
        \textbf{Total} & & & \textbf{7682} & \textbf{640}\\
        \bottomrule
    \end{tabular}
    \label{tab:data_info}
\end{table}

We ran the classification model BirdNET \cite{birdnet} on each file in the dataset, which partitions the data and classifies each 3 second clip. The model produced 67,163 detections in total, or approximately 56 hours, representing 8.7\% of the dataset. The data were refined further by selecting only those bird classes with at least 100 examples. We chose this threshold to ensure reliable estimation of per-class accuracy metrics and to allow meaningful training and validation of both the generative and classification models. For each of these species, we selected up to 500 of the most confidently identified examples (confidence over 0.5) by BirdNET. Our final filtered dataset has 8,248 audio clips involving 27 distinct bird species. A further 17,810 clips with confidence between 0.25 and 0.5 were included in the training and validation sets in later runs for a more generalised model, for a total of 26,058 audio clips.

The BirdNET pseudo-labelled dataset was used as a training and validation set, with a 90-10 split stratified by class to deal with the imbalanced distribution of the data. Table~\ref{tab:dist} shows a list of the classes together with their split in terms of training, validation and test sets by class. 

The test set consists solely of human-labelled audio samples, where experts annotated the data with NEAL \cite{neal} software with each annotation containing a confidence score. The annotated data are from the same sites as listed in Table~\ref{tab:data_info} but from files that are not included in the training or validation datasets. Keeping the test set distinct ensures that it remains independent, preventing overfitting and data leakage. After some remapping of class names to account for slight discrepancies with custom labels, and including only those with one of the 27 classes mentioned above, we arrived at a test set with 1,334 examples. These labels were in the form of bounding boxes and could be up to 15 seconds long. We thus only took the first 3 seconds of each label to match with BirdNET samples. Cropping the annotations thus has a chance of excluding parts of the vocalisations that occur later in the clips but the annotations always begin when the species is already clearly vocalising. After removing labels with labeller confidence less than $0.9$, the final test set had 825 examples. 

The test set distribution is different to both training and validation sets, with some of the 27 classes not being in the test set. We chose to keep those extra classes not in the test set in order to 1) give the generative models more data to learn from and 2) train more generalised, robust classifiers, i.e., models capable of handling changes in the data distribution and maintaining performance across varied conditions as would be required in noisy environments such as wind farms. Furthermore, due to the limited size of the test set, we decided not to include any real examples in the training or validation data and preserve the full amount as a human benchmark. During evaluation of the test set we ignored these missing classes when calculating the accuracy of predictions for the classifiers. 

\begin{table}[!ht]
\caption{\textbf{Distribution of species in the dataset.} The data were split via class-stratified random sample: 90\% for training, 10\% for validation, with human labels as test. The horizontal bars visualise the distributions. Note that some classes are missing in the test set because of the limited amount of human-verified samples. For example, Carrion Crow is quite rare in Ireland, and does not appear in the test set. Examination suggests that these pseudo-labels are likely to be Hooded Crow.}
    \centering
    \begin{tabular}{rlrrrlll}
  \hline
 & Common name & Training count & Validation count & Test count & \% train & \% val & \% test \\ 
  \hline
1 & Barn Swallow & 299 &  33 &   0 & \color{seabornBlue}{\rule{0.242cm}{6pt} 4} & \color{seabornOrange}{\rule{0.24cm}{6pt} 4} & \color{seabornGreen}{\rule{0cm}{6pt} 0} \\ 
  2 & Carrion Crow & 250 &  28 &   0 & \color{seabornBlue}{\rule{0.202cm}{6pt} 3.4} & \color{seabornOrange}{\rule{0.204cm}{6pt} 3.4} & \color{seabornGreen}{\rule{0cm}{6pt} 0} \\ 
  3 & Common Chaffinch & 195 &  22 & 112 & \color{seabornBlue}{\rule{0.158cm}{6pt} 2.6} & \color{seabornOrange}{\rule{0.16cm}{6pt} 2.7} & \color{seabornGreen}{\rule{0.339cm}{6pt} 13.6} \\ 
  4 & Common Wood-Pigeon & 250 &  28 &   3 & \color{seabornBlue}{\rule{0.202cm}{6pt} 3.4} & \color{seabornOrange}{\rule{0.204cm}{6pt} 3.4} & \color{seabornGreen}{\rule{0.009cm}{6pt} 0.4} \\ 
  5 & Dunnock & 450 &  50 &  11 & \color{seabornBlue}{\rule{0.364cm}{6pt} 6.1} & \color{seabornOrange}{\rule{0.364cm}{6pt} 6.1} & \color{seabornGreen}{\rule{0.033cm}{6pt} 1.3} \\ 
  6 & Eurasian Blackbird & 414 &  46 & 251 & \color{seabornBlue}{\rule{0.335cm}{6pt} 5.6} & \color{seabornOrange}{\rule{0.335cm}{6pt} 5.6} & \color{seabornGreen}{\rule{0.761cm}{6pt} 30.4} \\ 
  7 & Eurasian Blue Tit & 383 &  43 &   3 & \color{seabornBlue}{\rule{0.31cm}{6pt} 5.2} & \color{seabornOrange}{\rule{0.313cm}{6pt} 5.2} & \color{seabornGreen}{\rule{0.009cm}{6pt} 0.4} \\ 
  8 & Eurasian Linnet & 450 &  50 &  41 & \color{seabornBlue}{\rule{0.364cm}{6pt} 6.1} & \color{seabornOrange}{\rule{0.364cm}{6pt} 6.1} & \color{seabornGreen}{\rule{0.124cm}{6pt} 5} \\ 
  9 & Eurasian Magpie & 121 &  13 &   0 & \color{seabornBlue}{\rule{0.098cm}{6pt} 1.6} & \color{seabornOrange}{\rule{0.095cm}{6pt} 1.6} & \color{seabornGreen}{\rule{0cm}{6pt} 0} \\ 
  10 & Eurasian Skylark & 429 &  48 &   0 & \color{seabornBlue}{\rule{0.347cm}{6pt} 5.8} & \color{seabornOrange}{\rule{0.349cm}{6pt} 5.8} & \color{seabornGreen}{\rule{0cm}{6pt} 0} \\ 
  11 & Eurasian Wren & 450 &  50 &  96 & \color{seabornBlue}{\rule{0.364cm}{6pt} 6.1} & \color{seabornOrange}{\rule{0.364cm}{6pt} 6.1} & \color{seabornGreen}{\rule{0.291cm}{6pt} 11.6} \\ 
  12 & European Goldfinch & 379 &  42 &   6 & \color{seabornBlue}{\rule{0.306cm}{6pt} 5.1} & \color{seabornOrange}{\rule{0.305cm}{6pt} 5.1} & \color{seabornGreen}{\rule{0.018cm}{6pt} 0.7} \\ 
  13 & European Greenfinch &  99 &  11 &   0 & \color{seabornBlue}{\rule{0.08cm}{6pt} 1.3} & \color{seabornOrange}{\rule{0.08cm}{6pt} 1.3} & \color{seabornGreen}{\rule{0cm}{6pt} 0} \\ 
  14 & European Robin & 450 &  50 &  63 & \color{seabornBlue}{\rule{0.364cm}{6pt} 6.1} & \color{seabornOrange}{\rule{0.364cm}{6pt} 6.1} & \color{seabornGreen}{\rule{0.191cm}{6pt} 7.6} \\ 
  15 & European Starling &  95 &  10 &  31 & \color{seabornBlue}{\rule{0.077cm}{6pt} 1.3} & \color{seabornOrange}{\rule{0.073cm}{6pt} 1.2} & \color{seabornGreen}{\rule{0.094cm}{6pt} 3.8} \\ 
  16 & European Stonechat & 450 &  50 &   9 & \color{seabornBlue}{\rule{0.364cm}{6pt} 6.1} & \color{seabornOrange}{\rule{0.364cm}{6pt} 6.1} & \color{seabornGreen}{\rule{0.027cm}{6pt} 1.1} \\ 
  17 & Goldcrest & 121 &  13 &   3 & \color{seabornBlue}{\rule{0.098cm}{6pt} 1.6} & \color{seabornOrange}{\rule{0.095cm}{6pt} 1.6} & \color{seabornGreen}{\rule{0.009cm}{6pt} 0.4} \\ 
  18 & Great Tit & 123 &  14 &   4 & \color{seabornBlue}{\rule{0.099cm}{6pt} 1.7} & \color{seabornOrange}{\rule{0.102cm}{6pt} 1.7} & \color{seabornGreen}{\rule{0.012cm}{6pt} 0.5} \\ 
  19 & Hooded Crow & 190 &  21 &  18 & \color{seabornBlue}{\rule{0.154cm}{6pt} 2.6} & \color{seabornOrange}{\rule{0.153cm}{6pt} 2.5} & \color{seabornGreen}{\rule{0.055cm}{6pt} 2.2} \\ 
  20 & Meadow Pipit & 450 &  50 &  65 & \color{seabornBlue}{\rule{0.364cm}{6pt} 6.1} & \color{seabornOrange}{\rule{0.364cm}{6pt} 6.1} & \color{seabornGreen}{\rule{0.197cm}{6pt} 7.9} \\ 
  21 & Redwing &  90 &  10 &   0 & \color{seabornBlue}{\rule{0.073cm}{6pt} 1.2} & \color{seabornOrange}{\rule{0.073cm}{6pt} 1.2} & \color{seabornGreen}{\rule{0cm}{6pt} 0} \\ 
  22 & Rook & 450 &  50 &  52 & \color{seabornBlue}{\rule{0.364cm}{6pt} 6.1} & \color{seabornOrange}{\rule{0.364cm}{6pt} 6.1} & \color{seabornGreen}{\rule{0.158cm}{6pt} 6.3} \\ 
  23 & Sedge Warbler & 108 &  12 &   1 & \color{seabornBlue}{\rule{0.087cm}{6pt} 1.5} & \color{seabornOrange}{\rule{0.087cm}{6pt} 1.5} & \color{seabornGreen}{\rule{0.003cm}{6pt} 0.1} \\ 
  24 & Spotted Flycatcher & 125 &  14 &   0 & \color{seabornBlue}{\rule{0.101cm}{6pt} 1.7} & \color{seabornOrange}{\rule{0.102cm}{6pt} 1.7} & \color{seabornGreen}{\rule{0cm}{6pt} 0} \\ 
  25 & White Wagtail & 232 &  26 &   5 & \color{seabornBlue}{\rule{0.188cm}{6pt} 3.1} & \color{seabornOrange}{\rule{0.189cm}{6pt} 3.2} & \color{seabornGreen}{\rule{0.015cm}{6pt} 0.6} \\ 
  26 & Willow Warbler & 119 &  13 &   0 & \color{seabornBlue}{\rule{0.096cm}{6pt} 1.6} & \color{seabornOrange}{\rule{0.095cm}{6pt} 1.6} & \color{seabornGreen}{\rule{0cm}{6pt} 0} \\ 
  27 & Yellowhammer & 251 &  28 &  51 & \color{seabornBlue}{\rule{0.203cm}{6pt} 3.4} & \color{seabornOrange}{\rule{0.204cm}{6pt} 3.4} & \color{seabornGreen}{\rule{0.155cm}{6pt} 6.2} \\ 
   \hline
\end{tabular}
    \label{tab:dist}

\end{table}

\section{Methods}\label{ch:methods}

Our study employs a multi-step approach to generate and evaluate the quality of spectrograms of bird sounds. We begin this section by describing the preprocessing steps involved converting the raw audio into mel spectrograms. We then give background on generative models used, how controllable generation allows for generating (labelled) samples of a specific class, and the differences between the reference models and ours. We assess the quality of synthetic images to decide on the better generative model to use as a data augmentation method. Finally, we give details on the equipment used in the generative and classification model training processes.

\subsection{Preprocessing}

In order to exploit a variety of image classification and generation techniques, the audio samples were converted into a visual format as per standard practise in the field \cite{Stowell2022-vh}. The audio recordings were transformed into a visual representation known as the \emph{mel spectrogram}. The methods we use require square images to work in the existing framework so we aim to transform the \verb|.wav| files into spectrograms that are approximately square. Mel spectrograms offer several advantages over traditional linear spectrograms; being an effective representation of time-frequency dynamics present in animal vocalisations, offering reduced dimensionality compared to raw waveforms, and aligning more closely with human auditory perception \cite{10.1121/1.1915893}. 

The spectrogram is computed from the raw waveform using \verb|librosa.feature.melspectrogram| \cite{librosa} and parameters: 256 mel filters (\verb|n_mels|), a Fast Fourier Transform (FFT) size of 512 (\verb|nfft|), and a \verb|hop_length| equal to 75\% of the \verb|nfft| value. The transformation is carried out on 3-second inputs with a sampling rate of 48kHz, consistent with BirdNET. The extra width in the mel spectrogram is then cropped to 256 wide and the resulting square matrix is converted from its amplitude squared to dB, which involves a \verb|log| transform. The resulting spectrogram is then saved using \verb|matplotlib.pyplot.imsave| \cite{matplotlib} to give a $256\times256$ single-channel (greyscale) image with pixel values from 0-255. Alternatively, the spectrograms could have been stored directly as floating-point tensors (e.g., \verb|.pt| or \verb|.npy| files), which would preserve full numerical precision. However, storing them as \verb|.png| images offered several practical advantages: it facilitated rapid visual inspection and quality assessment of thousands of spectrograms during preprocessing, enabled seamless use of standard \verb|torchvision| image-loading and augmentation pipelines, and aligns with widespread practice in the audio representation-learning literature. When the images are read in for training models, a torch transform normalises each spectrogram's pixel values to the range -1 to 1.

Although numerous data-augmentation techniques exist for audio classification tasks (applicable either directly to waveforms or to spectrogram images), we decided not to apply them as additional preprocessing layers for this paper. Preliminary experiments with waveform-level techniques—including UNet-based denoising, mixup with other samples from the same dataset, and random time shifting—yielded no consistent improvement in validation performance. We therefore chose to use entirely generative methods as data augmentation techniques.

\subsection{Generating Artificial Spectrograms}

\subsubsection{Generative Adversarial Networks (GAN)}

Generative Adversarial Networks (GANs) are a class of machine learning model consisting of two neural networks, that is, a generator and a discriminator, that compete in a two-player minimax game \cite{goodfellow2014gan}. The generator takes in a latent (or noise) vector $z$ and outputs synthetic data resembling real samples, while the discriminator attempts to differentiate between the real and synthetic data. Both models start out with randomly initialised weights and improve in tandem. As the generator improves, it becomes better at fooling the discriminator, leading to increasingly realistic data outputs, i.e. minimising the loss function conveying the difference in distribution between the real and synthetic data. Conversely, as the discriminator improves, it becomes better at distinguishing real from synthetic data, i.e. maximising this same loss function. At the end of the training process, the generator is used for generating synthetic data, while the discriminator can be discarded or used to determine the quality of the generated images. GANs have gained widespread attention for their ability to generate high-quality images \cite{https://doi.org/10.1016/j.ecoinf.2023.102321, https://doi.org/10.1016/j.ifacol.2019.12.406}, audio \cite{10.1109/LSP.2022.3150258, 10.1109/ICIVC55077.2022.9886154}, and video \cite{https://doi.org/10.1016/j.neunet.2020.09.016, 10.1007/s11263-020-01333-y} content as well as biological sequencing \cite{https://doi.org/10.3390/biology12060854}, camouflage design \cite{https://doi.org/10.1111/2041-210X.13334}, and classifying bats using radar \cite{https://doi.org/10.1111/2041-210X.14125}.

Auxiliary Classifier GANs (ACGAN) build on the original GAN architecture by incorporating class information to the framework \cite{odena2017acgan}. The generator's latent vector $z$ is now a concatenation of a one-hot encoded vector of class information $y$ and a noise vector. The discriminator is given the  additional task of predicting class labels for both real and generated data, as with a standard image classifier. In an ACGAN, the discriminator outputs both the probability that the input is real or fake, as in the standard GAN, as well as the probability of the image coming from each class. The addition of an auxiliary task has two key benefits: it allows ACGANs to create more realistic and diverse images as the classifier component of the discriminator's loss is backpropagated through both models \cite{sadat2024cadsunleashingdiversitydiffusion}, and the generator is now capable of synthesising fake examples with explicit class information guiding the model. The latter is known as controllable generation. 

By incorporating species/class information into the training process, ACGANs allow for more fine-tuned generation of synthetic examples. These samples can later be filtered (if necessary) to keep just the highest quality samples, as determined by the discriminator, which can then be iteratively added to the training data. These larger datasets, depending on their design, can have increased diversity or help address class imbalance, enhancing the robustness of the training process. The ACGAN extension has been applied to acoustic classification \cite{10.1121/10.0005202,dracgan} to generate realistic spectrograms to supplement and enhance the existing datasets, which then improves classification accuracy of the models trained on these datasets.

\begin{figure}[!ht]
    \centering
    \includegraphics[width=0.99\textwidth]{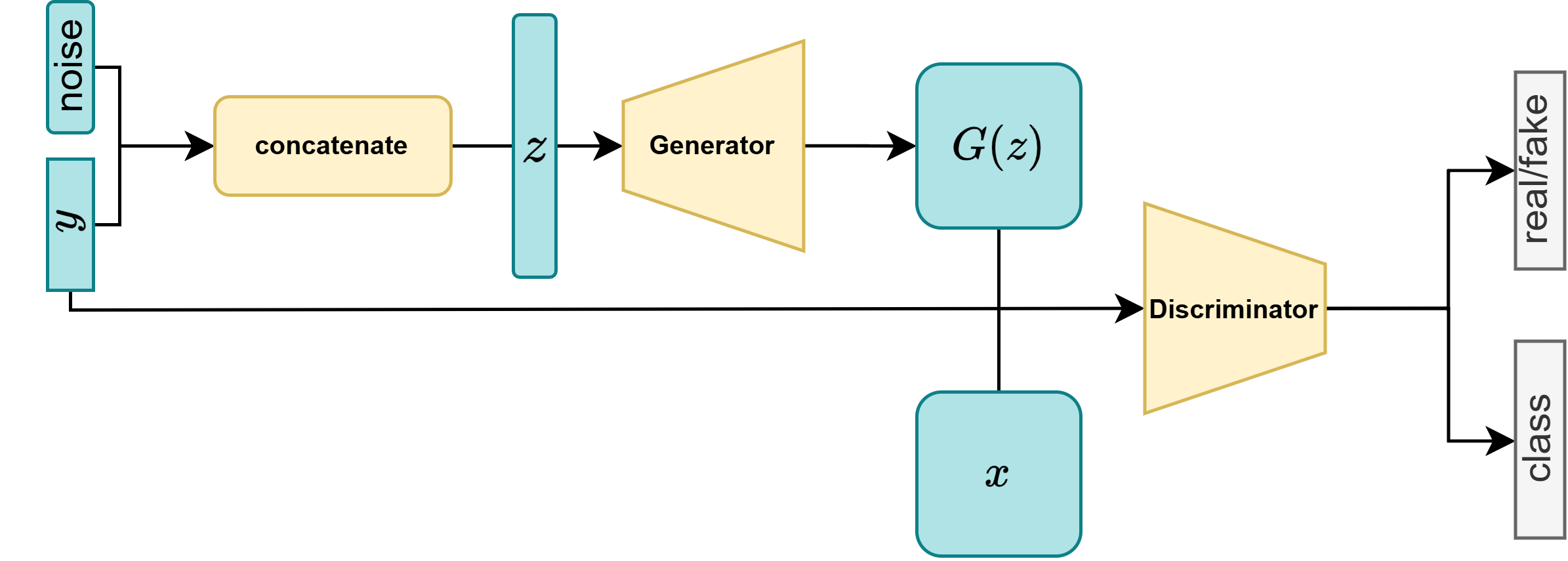}
    \caption{\textbf{Auxiliary Classifier GAN (ACGAN) architecture.} For training, real images $x$ and their (one-hot encoded) labels $y$ are given, and sample noise vectors and $y$ are concatenated into latent vectors $z$. The Generator $G$ generates images $G(z)$ which come from the same distribution as $x$. Both $x$ and $G(z)$ are passed to a Discriminator $D$ which predicts 1) whether images are real or fake, and 2) their classes. A \emph{two-player minimax game} arises, where $G$ learns to create images to fool $D$ and $D$ learns to better tell $x$ and $G(z)$ apart even as $G$ improves. Images are then generated using the generator as above for subsequent sampling.
    }
    \label{fig:gan_architecture}
\end{figure}

We trained an ACGAN to generate mel spectrograms for supplementing our dataset. We wrote the training script and model definitions in PyTorch. The Generator and Discriminator block architectures were based on \cite{dracgan}. We then modify the setup as follows:
\begin{itemize}
    \item single channel $256\times256$ images were used instead of three-channel images, as spectrograms, being time-frequency representations of sound, are inherently greyscale and do not require the RGB colour structure.
    \item an open source PyTorch implementation that can be adapted for future use. The training loop was based on code from \cite{Wu2018clvrai}.
    \item using negative log likelihood and log-softmax outputs rather than categorical cross-entropy and raw logits as the discriminator auxiliary loss for more training stability. We found that training on data with significant background noise required this step.
\end{itemize}
The architecture of the ACGAN we use is shown in  Figure~\ref{fig:gan_architecture}.

\subsubsection{Denoising Diffusion Probabalistic Models (DDPM)}

\begin{figure}[!ht]
    \centering
    \includegraphics[width=0.99\textwidth]{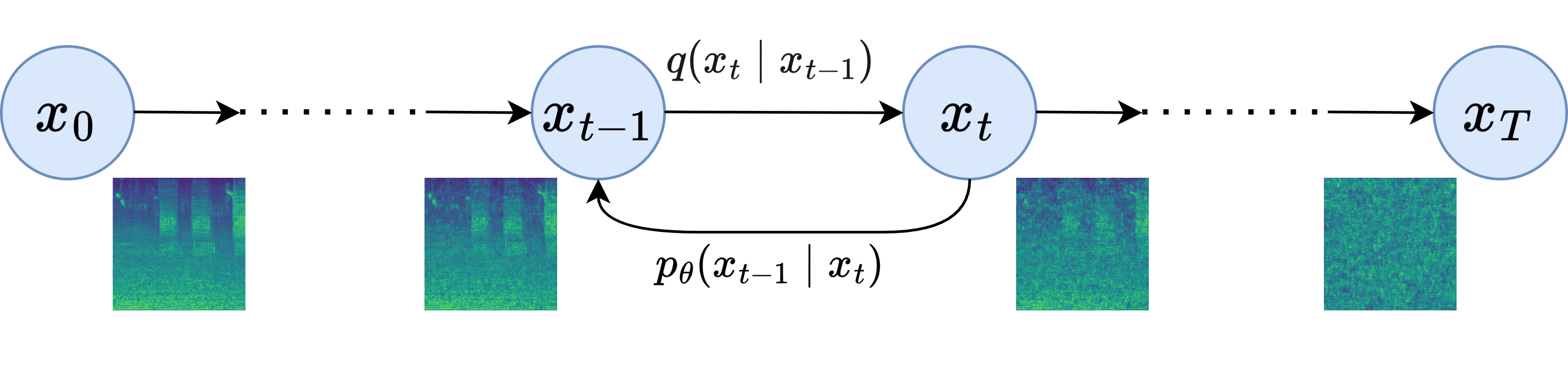}
    \caption{\textbf{Denoising UNet from diffusion model.} We have a defined forward process $q(x_t\mid x_{t-1})$ to add noise to an image $x_{t-1}$ to create an image $x_t$. The reverse process $p_\theta(x_{t-1}\mid x_t)$ is performed by the UNet with parameters $\theta$.}
    \label{fig:add_noise}
\end{figure}

Diffusion models are another class of generative AI model that generate data by gradually transforming noise into structured samples. The models utilise a two-phase process, a forward diffusion process (adding noise to the data) and a reverse diffusion process, which undoes the forward process \cite{sohldickstein2015diffusion}. In the forward process, Gaussian noise is progressively added, $t$ times, to the data $x_0$ to produce a somewhat degraded image $x_t$, up to a maximum degradation $x_T$ for some $T$ which would be pure Gaussian noise. The reverse diffusion process uses a UNet model, a convolutional neural network architecture, to predict the noise that was added and produce a reconstruction $\tilde{x}_0$ of the original image from $x_t$, i.e. $\tilde{x}_0=x_t - \text{UNet}(x_t)$. During the image generation (inference) phase, some pure Gaussian noise, $x_T$, is generated and the predicted noise is successively subtracted one step at a time, to generate $\tilde{x}_0$, the final synthetic image. Figure~\ref{fig:add_noise} shows a schematic of the DDPM process. 

DDPMs are a specific type of diffusion model that formalises this noise-reduction process using a probabilistic framework \cite{https://arxiv.org/abs/2006.11239}. The training loop involves maximising the likelihood of reconstructing the data from its noisy counterpart. The timestep $t$ is also passed to the model allowing the model to learn the conditional probabilities at each step of the diffusion process (from $0$ to $T$), which enables it to denoise the data step by step in reverse. The UNet takes in the noisy image and outputs the predicted noise component, as in the above equation for $\tilde{x}_0$. The loss then compares actual and predicted noise. The model's ability to successively remove noise from input images results in high-fidelity images comparable or superior in quality to other generative models such as GANs. Diffusion models for data generation have seen wide application in medical imaging \cite{https://doi.org/10.48550/arXiv.2210.04133, https://doi.org/10.1016/j.imu.2024.101575}, music synthesis \cite{Smith2024teticio}, molecule design \cite{weiss2023guided}, and filling in missing or degraded parts of an image \cite{lugmayr2022repaintinpaintingusingdenoising}.

Stable Diffusion models, originally known as Latent Diffusion Models, are similar in design to DDPMs, with the extra step of first encoding the images into a latent space, for example using an autoencoder \cite{rombach2022highresolutionimagesynthesislatent}. The images (in what is called pixel space) are transformed into the latent space, where the forward and reverse (probabalistic) diffusion processes are carried out, before being decoded back to the original image size in pixel space. Since the UNet is the more computationally intensive model, and it is run many (up to $T$) times during generation, it is much more efficient to carry out this step in the latent space than pixel space, allowing for faster generation and high-resolution outputs.

\begin{figure}[!ht]
    \centering
    \includegraphics[width=0.99\textwidth]{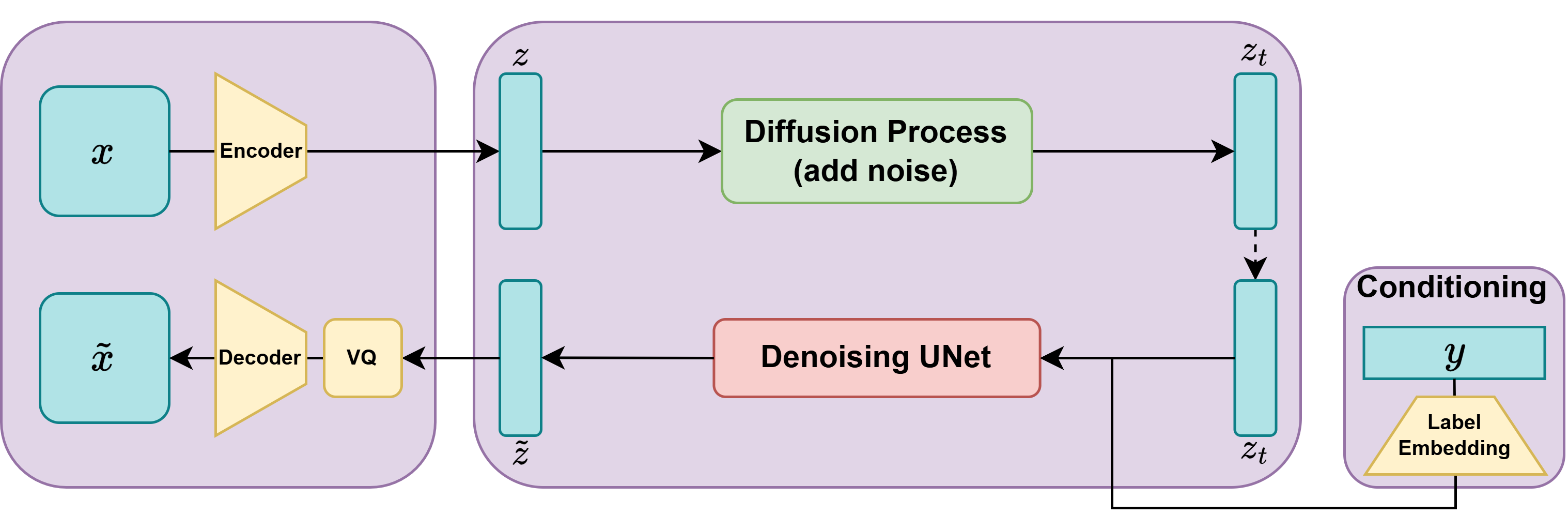}
    \caption{\textbf{Conditional Latent Diffusion architecture.}
For training, an image $x$ is encoded to a latent $z$ using the VQAE encoder. A random timepoint $t$ from $1,\dots,T$ is taken. Noise is added to $z$ for $t$ steps giving $z_t$. The UNet, with inputs $z_t$, $t$ and $y$,  outputs predicted noise and the loss compares the predicted noise with the actual noise added to $z$. For sampling, a latent $z$ and class $y$ are generated/given, either pure noise or a partially noised image. For $T$ steps, the UNet runs with $z$, $t=1$ and $y$ as inputs and the output subtracted from $z$ itself, updating $z$. Once denoised $T$ times, this $\tilde{z}$ is vector quantised, decoded to $\tilde{x}$ and saved.}
    \label{fig:diffusion_architecture}
\end{figure}

We employ a stable diffusion model and embed species information for training and generation, as we did with ACGAN. We based our stable diffusion model on the conditional DDPM created by \cite{Capelle2023tcapelle} and modify it as follows: 
\begin{itemize}
    \item single channel input images were used instead of three-channel images, as spectrograms are inherently greyscale and do not require the RGB colour structure.
    \item work with a lower dimensional latent space with tensors of shape $64\times64\times4$ to train the UNet rather than the $256\times256$ images themselves, making it a stable diffusion model,
    \item using a Vector Quantised-Variational AutoEncoder (VQVAE) to encode to and decode from latent space \cite{https://arxiv.org/abs/1711.00937}, which has the benefits of producing a discrete latent representation, more learnt behavior in terms of the prior and avoiding posterior collapse.
\end{itemize}
Figure~\ref{fig:diffusion_architecture} shows our model architecture diagram (based on \cite{rombach2022highresolutionimagesynthesislatent}).

\subsection{Evaluating the Quality of the Generated Spectrograms}

We evaluated the quality of the synthetic images generated using two metrics: the Inception Score (IS) \cite{salimans2016improvedtechniquestraininggans} and the Fréchet Inception Distance (FID) \cite{NIPS2017_8a1d6947}. IS measures the diversity and clarity of the generated images, ranging from zero (worst quality) to $N$ (best quality), where $N$ is the number of classes. FID assesses the similarity between the generated and real images and ranges from 0 to infinity, with a lower FID indicating closer resemblance to the actual dataset. These metrics were used to compare generated samples from both the GAN and Diffusion models, to evaluate how effectively each set of synthetic examples replicate the characteristics of actual spectrograms.

While both image quality metrics have previously been used to measure the quality of spectrograms \cite{dracgan}, there have been issues raised with measures like Inception Score \cite{barratt2018noteinceptionscore}. These stem mainly from the fact that its predictions come from the Inception V3 model, which was trained in a different modality and set of classes (in our case ImageNet rather than spectrograms of bird calls). However, it is primarily used as a heuristic to sanity check the quality of training samples after what can be a lengthy training and generation process, rather than as an all-encompassing benchmark. The diffusion samples were more convincing than the GAN samples and so human judgment would have prevailed in either case.

To obtain a more appropriate assessment of audio quality, we converted the generated spectrograms back to audio waveforms using the Griffin-Lim algorithm \cite{griffin1984signal}. We then evaluated the resulting signals with the Fréchet Audio Distance (FAD) \cite{kilgour2019frchet,gudgud96fad}. FAD functions like FID, but with embeddings coming from an audio model. We chose embeddings from the CLAP audio encoder \cite{elizalde2023clap}, which is trained contrastively on diverse audio-text pairs and thus well-suited for general-purpose audio evaluation. Lower values of FAD suggest the generated audio samples more closely align with the real audio dataset in the embedding space.

\subsection{Experiment Environment}

The hardware environment for running our experiments includes a Unix Server, accessed via a web browser, with 566 GB of memory, 
an Intel(R) Xeon(R) Silver 4214 CPU @ 2.20GHz with 12 cores and 24 threads, 
and an NVIDIA Tesla T4 GPU with 15 GiB memory. 
The operating system is Ubuntu 22.04.4 LTS.
The programming environment consists of Visual Studio Code version 1.93.1 \cite{vscode} for prototyping, training and post-processing. Within this, we used CUDA Version: 12.2 \cite{cuda}, Python 3.10.12 \cite{python3.10.12}, and PyTorch 2.0.1+cu117 \cite{pytorch2.0.1}.

\section{Experimental Design and Results}\label{ch:results}

\begin{figure}[!ht]
    \centering
    \setlength{\tabcolsep}{1pt} 
    \renewcommand{\arraystretch}{1.5} 
    \includegraphics[width=0.99\textwidth]{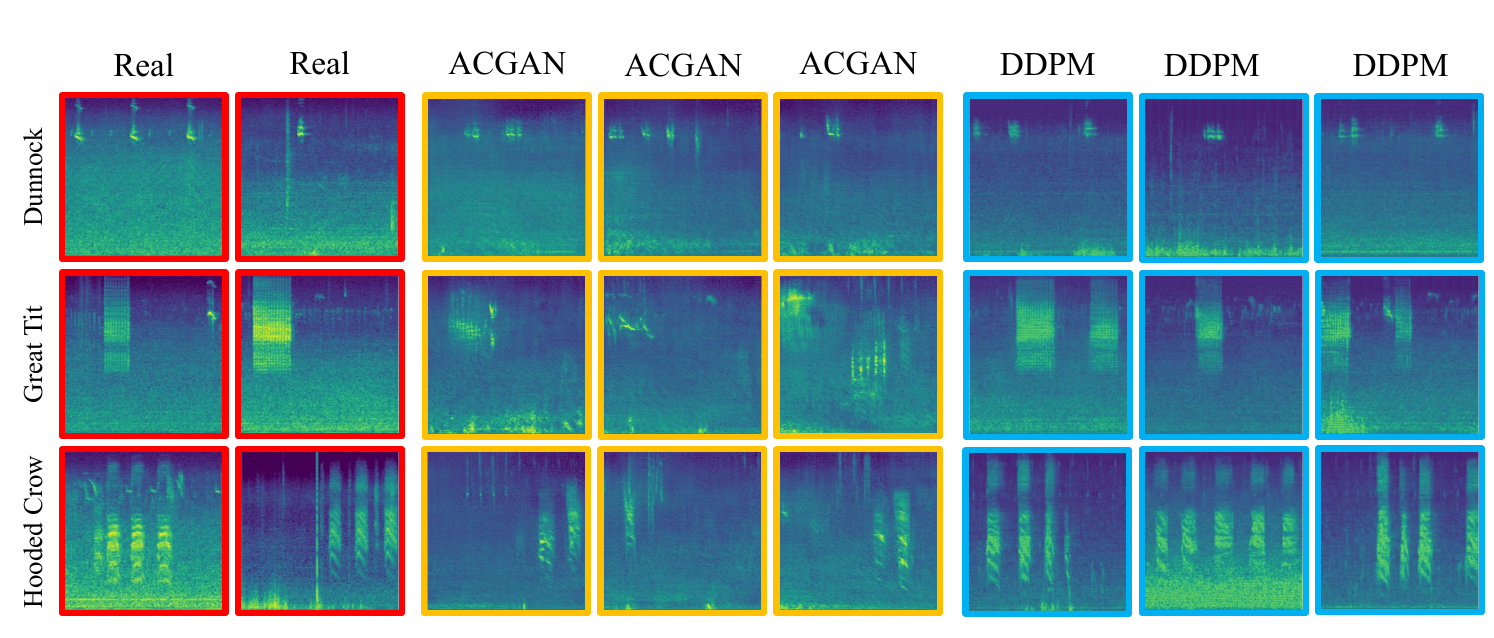}
    \caption{\textbf{Real and Synthetic Spectrograms.} The first two (leftmost) red-outlined images in each row are real spectrograms. The next three orange-outlined images are generated by the ACGAN, while the last three blue-outlined images are synthetic samples generated using the DDPM. The $x$-axis of each spectrogram represents time, ranging from 0-3 seconds. The $y$-axis represents frequency, ranging from 0 to 12kHz on the mel scale. Brighter colors (using the viridis colour palette) indicate higher energy or loudness. The ACGAN-generated Dunnock and Hooded Crow samples do show some species-specific features, but the Great Tit samples are poor imitations. The DDPM-generated samples show close similarity to the real examples in red but also have variety. The Great Tit examples generated by DDPM are clearly more convincing than the ACGAN samples.}
    \label{tab:real_synth}
\end{figure}

Both generative models were trained for 200 epochs with square, greyscale images of size 
$256\times256\times1$ (see training results in Figure~\ref{fig:losses}). 
After training, both models were set to generate a random collection of images with equal weighting across classes, a total of  $250 \times 27=6,750$ synthetic images from each model. Figure~\ref{tab:real_synth} shows some example images synthesised by each generative model. While the ACGAN-generated samples show some clear learning of distinct features for each of the species, they often overly focus on the background noise, with some artefacting or undesirable feature exaggeration observed. Conversely, the DDPM-generated spectrograms are not only convincing but diverse, and appear higher quality than the GAN samples. The structure of the background noise also looks reasonable, with little artefacting. Whilst visually appearing better than the ACGAN samples, next we provide more numerical comparisons of the performance of the generated spectrograms.

Image quality scores support this observation, with DDPM samples showing a 7\% increase in IS and 51\% decrease in FID score relative to ACGAN samples (Table~\ref{tab:quality_metrics}). Furthermore, DDPM samples also perform better on the audio quality metric FAD, showing an 8.4\% lower score than ACGAN. The scores reinforce our informal visual inspection. We therefore decided to thoroughly evaluate the the utility of supplementing our training data with DDPM samples for training acoustic classifiers.

\begin{table}[!ht]
    \caption{\textbf{Synthetic data quality metrics.} The baseline IS, FID and FAD scores were calculated on the real data and thus serve as a target value for those created by ACGAN and DDPM. Higher IS scores and lower FID and FAD scores indicate better performance. Note that the FID and FAD scores comparing the dataset with itself are both zero.}
    \centering
    \begin{tabular}{L{3cm}R{2cm}R{2cm}R{2cm}}
        \hline
        Model & IS & FID & FAD\\ \hline
        Baseline/target & 2.00 & 0.00 & 0.00\\
        ACGAN & 1.62 & 64.97 &  1.07 \\ 
        DDPM &  \textbf{1.73} & \textbf{32.05} & \textbf{0.98}\\ \bottomrule
    \end{tabular}
    \label{tab:quality_metrics}
\end{table} 

Next, we use our generated samples to determine whether they can assist in improving the classification performance of the model. We set up our experimental datasets as follows: The training data was supplemented with \verb|[0,50,100,150,200,250]| DDPM-generated samples per class, providing up to 6,750 more images (an $\sim$81\% increase in data set size) to learn structures in the spectrograms for distinguishing bird species. The validation and test datasets remained unchanged to prevent data leakage from using synthetic examples derived from the training data.

For each dataset combining real and (possibly no) synthetic examples, a selection of off-the-shelf classification models were chosen: MobileNetV2 \cite{sandler2019mobilenetv2invertedresidualslinear}, ResNet18 \cite{he2015deepresiduallearningimage}, and VGG16 \cite{simonyan2015deepconvolutionalnetworkslargescale}. These models were trained via transfer learning, where early layers were kept frozen but with the last few layers being trainable. The weights were loaded from a version pre-trained on ImageNet, a large scale computer vision benchmark dataset \cite{5206848}. The final layer of each was overwritten to match the number of classes for this dataset (27) and whose weights were randomly initialised. To provide an alternative to the pre-trained models, we also included a small custom Convolutional Neural Network (CNN) trained from scratch (see Table~\ref{tab:custom_architecture}). Finally, once all four of these models were trained (that is, MobileNetV2, ResNet18, VGG16, and our custom model), an ensemble model \cite{zhou2012ensemble} was constructed by taking the concatenated $108$-dimensional probability vectors ($27$ classes $\times \ 4$ models) produced by the individual classifiers and feeding them into a single trainable linear layer that outputs the final 27-class distribution. The linear layer was trained on the same training set while keeping the base models fixed. A comparison of the model sizes is given in Table~\ref{tab:model_metrics}.

\begin{figure}[!ht]
    \centering
    \includegraphics[width=\textwidth]{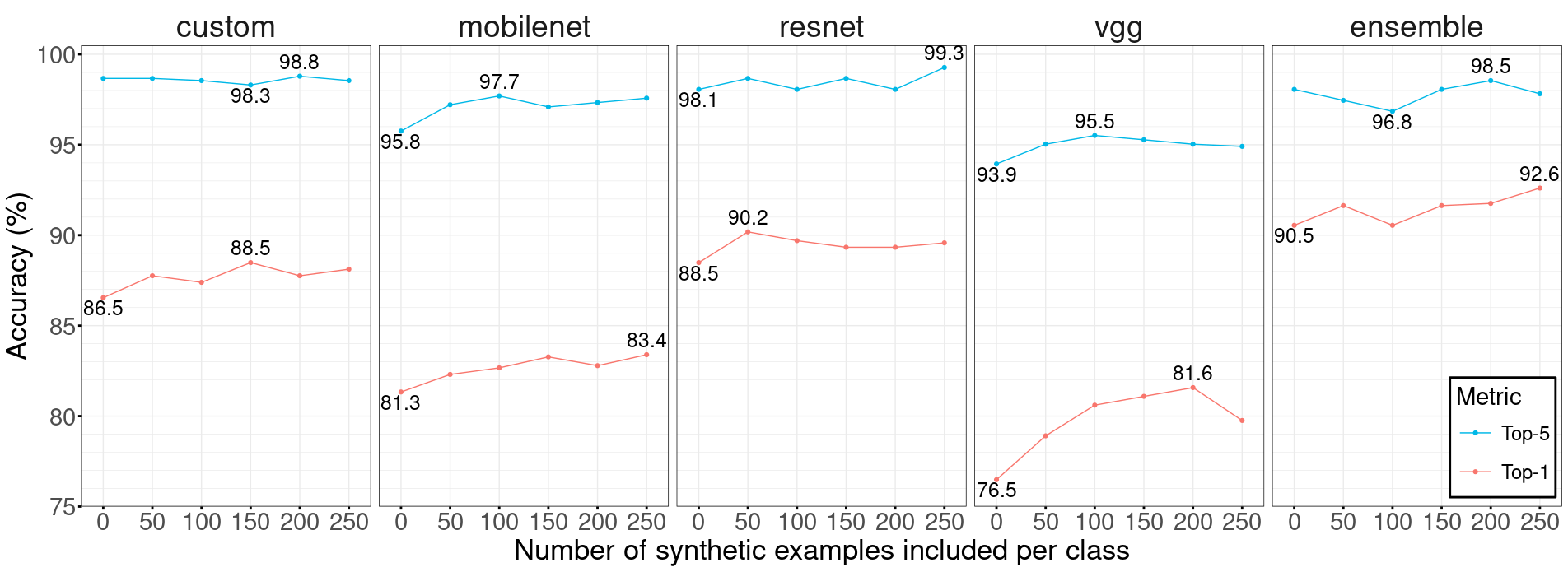}
    \caption{\textbf{Classification Accuracy Results.} Top-1 (red line) and Top-5 (blue line) validation accuracy (y-axis) with varying synthetic data per class (x-axis). All models show a positive trend in accuracy improvement when including synthetic examples, particularly VGG. While the ResNet model's accuracy decreases beyond 50 additional synthetic samples per class, these values are still better than with only real data. 
    }

    \label{fig:val_acc}
\end{figure}

Each of the five above models used the same training hyperparameters: Adam optimiser, learning rate of 0.001, a batch size of 32, and 50 epochs. For the evaluation of these classifiers, a set of metrics was employed to assess their performance comprehensively: accuracy, precision, recall, F1 score, and top-5 accuracy. Figure~\ref{fig:val_acc} shows the best validation accuracy for each training process across different models and levels of synthetic data added. While there is an upward trend seen in classification performance by adding synthetic labelled samples, the validation set used also came from BirdNET pseudo-labels.

A better test of each model’s robustness uses human-verified samples. Although the ensemble model achieved the highest validation accuracy, it lagged behind BirdNET on the test set — likely because training relied on only highly confident BirdNET predictions. To build a more robust model, we retrained the ensemble (with and without synthetic examples) using knowledge distillation, where the model learns BirdNET’s output distributions rather than one-hot labels. This approach should better transfer BirdNET's insights as a state-of-the-art bird call classifier. 

As mentioned in Section \ref{ch:materials}, a larger dataset of 26,058 samples was used, again with a 90-10 split for training and validation. This was done in order for the model to learn the more difficult examples and better approximate BirdNET's distribution than only its most confident outputs. However, including less confident pseudo-labels to enhance generalizability risks incorporating a non-negligible number of incorrect labels, a problem potentially exacerbated by background noise from the wind farm data. 

\begin{figure}[!ht]
    \centering
    \includegraphics[width=\textwidth]{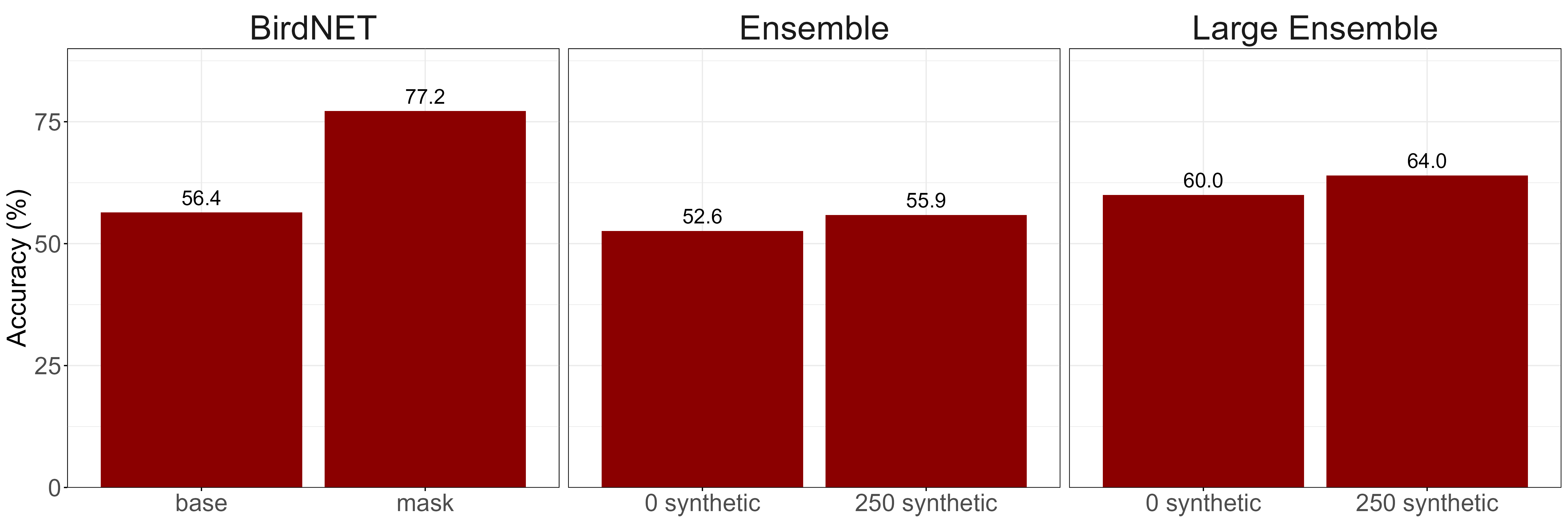}
    \caption{\textbf{Test Set Evaluation.} The bars represent the classification accuracy of models on human-labeled test data. Models include: BirdNET-base (comparing against the model's single class prediction), BirdNET-mask (using the maximum probability from the 27 indices in the BirdNET embeddings corresponding to our bird species of interest, rather than all classes), Ensemble (the ensemble of four models above retrained with knowledge distillation, using 0 or 250 synthetic samples per class), and Large Ensemble (ensemble retrained on a larger dataset including less confident BirdNET predictions, again with knowledge distillation and 0 or 250 synthetic samples per class). The Ensemble model with both real and synthetic samples improves accuracy by 3.3\% compared to real samples alone but underperforms BirdNET-base and BirdNET-mask. The Large Ensemble model achieves 60.0\% and 64.0\% accuracy with real and real plus 250 synthetic samples, respectively, surpassing the BirdNET-base but not BirdNET-mask. Further improvements would require more diverse data and higher-quality samples (e.g., including human-labelled data in training and validation).
    }
    \label{fig:test_acc}
\end{figure}

To train using knowledge distillation, the logits from the embeddings were converted into probabilities using the softmax function, with a temperature $T=3$, Kullback-Leibler (KL) divergence was used as the distillation loss and cross-entropy as the loss for the hard labels. The losses were combined with a weight of $\alpha=0.7$ for the distillation loss and $1 - \alpha = 0.3$ for the cross-entropy loss. Figure~\ref{fig:test_acc} shows that, while the models trained on BirdNET’s predictions from the real data alone do give reasonable performance on the test data somewhat, the models trained with both real and synthetic data outperform the ones trained on real data alone. 

\section{Discussion}\label{ch:discussion}
The synthetic images generated by Stable Diffusion appeared more realistic than those produced by ACGAN. Training off-the-shelf models on a dataset combining the original data with DDPM-generated samples yielded validation accuracy consistently higher than using real data alone. This suggests that increasing data quantity and quality can be as effective as, or more so than, model-centric improvements or domain-specific features engineering \cite{sutton2019bitter}.


The classifiers employed included MobileNetV2, ResNet18, VGG16 with ImageNet-pretrained weights and the last few layers unfrozen, and a small custom CNN trained from scratch. An ensemble model was trained, combining outputs from these four models and adding a final linear layer. The ensemble model, augmented by training with 250 extra synthetic examples added per species, achieved the highest accuracy overall on the validation set of 91.3\%.

Performance was evaluated on a hold-out test set with human-verified labels from the same wind farm sites as the pseudo-labelled data, comparing six setups: BirdNET, BirdNET with non-target (not in the 27 chosen) classes masked, an ensemble of four models trained on real and real plus synthetic data, and the same with a larger, less confident pseudo-labelled dataset with or without synthetic data. These setups achieved accuracies of 56.4\%, 77.2\%, 52.6\%, 55.9\%, 60.0\% and 64.0\%, respectively. Our results suggest that incorporating high-quality synthetic spectrograms into training data improves bird song classification performance on expert-labelled data.

The largest performance gain came from expanding the dataset to include more real samples (7.4\%), but required a $3.2\times$ increase in the dataset size. Augmenting both the initial and larger real datasets with synthetic examples further improved performance to 3.3\% and 4.0\%, respectively. However, this only required adding 82\% and 26\% more samples to the datasets, respectively. Although real data scaling yielded the largest individual gains in classification performance, adding synthetic samples provided meaningful additional improvements  more efficiently in terms of data volume, collection effort, and training time.

There still exists a notable gap between BirdNET's performance (77.2\% for BirdNET-mask) and the best ensemble classifier (64.0\%) on the hold-out test set. Several factors may contribute to this discrepancy. First, our dataset of 26,000 spectrograms corresponds to less than 24 hours in total, which may not be enough to generalise quite well enough even to a handful of habitats on Irish wind farms. BirdNET's was trained on a vast global dataset and likely captures a broader range of acoustic features from a wide variety of habitats and variations in bird vocalizations, enabling it to generalize better to the noisy wind farm environment. Second, the reliance on entirely pseudo-labelled data, processed by BirdNET itself, especially in the larger, less confident dataset, may introduce noise in the training process, as a non-negligible portion of these labels could be incorrect, as noted earlier. Additionally, while synthetic spectrograms improved performance, their quality and diversity may not fully replicate real-world variations, limiting the classifiers' ability to match BirdNET.

We have identified some further limitations with our approach, which we list below to support future work:
\begin{itemize}
\item \textbf{Automatic data pruning.} Several generated samples were noticeably distorted or contained artifacts, reducing their realism. A pruning method of removing such poor examples at scale would improve the quality metrics further and likely lead to more relevant learning.
\item \textbf{Pseudo-label bias.} Both the generative and classification models were trained and tuned by BirdNET pseudo-labels, with the embeddings used for the knowledge distillation of the final embedding models. This poses a risk of inheriting BirdNET's biases and systematic errors, especially given it may itself have been trained on very little data from a domain close to our noisy environment (Irish wind farms).
\item \textbf{Computational overhead.} The DDPM generation phase is slow, requiring approximately 1,000 denoising steps per image batch. Recent work \cite{https://doi.org/10.1111/2041-210X.14239}, using a similar VQ-VAE architecture, has improved efficiency in synthetic bird sound generation.

\item \textbf{More controllable generation.} Generative AI can be guided by adjusting specific input parameters. While species labels are effective here, other metatata such as weather, call type or time information could improve the variety of samples and align better with users' needs. Some Stable Diffusion papers \cite{10754260} have used an inpainting mask to direct the image generation process to specific areas in the spectrogram image.
\end{itemize}


The audio classification and generation approach developed for bird species can be adapted for other taxa with distinct vocalizations, such as bats, anurans, or marine mammals. For species with high-frequency calls, such as bats or certain insects, the preprocessing steps should be adjusted to target the relevant frequency range common to all species being classified, enhancing clarity of the spectrogram for deep learning models. For both generative and classification models, the question of whether to train a single model for multiple taxa or a separate models per taxon is worth exploring. By reducing reliance on human-labelled data, this method could support broader conservation efforts through enhanced species classification.

\section{Conclusion}\label{ch:conclusion}

Incorporating generative AI into bioacoustic pipelines can significantly enhance classification accuracy on domain-specific data, offering ecologists a valuable tool to supplement datasets constrained by data collection logistics and costs. This study demonstrates that augmenting bird species audio classification tasks with high-quality synthetic data generated using a Stable Diffusion model, improves performance in noisy environments compared to relying solely on the original data. Adding up to 250 DDPM-generated samples per class improved classification accuracy by 3.3\% for the ensemble model and up to 64.0\% for the large ensemble model on human-labeled test data, though it did not surpass BirdNET's accuracy. The DDPM-generated spectrograms were highly convincing, achieving a 7\% higher Inception Score (1.73 vs. 1.62),  a 51\% lower FID score (32.05 vs. 64.97) and 8\% lower FAD score (0.98 vs. 1.07), and underscore the potential of synthetic data augmentation over traditional approaches like model expansion or fine-tuning and so warrants further research. We foresee broader applications of synthetic generation of bird sounds, emphasizing the transformative impact of generative AI on ecological research.

\subsection*{Author Contributions}
AG, ID and AP conceived the ideas and designed methodology; AG and EK collected the data; All authors analysed the data; AG wrote the code and made the data available; AG, ID and AP led the writing of the manuscript. All authors contributed critically to the drafts and gave final approval for publication.

\subsection*{Acknowledgments}
 This paper comprises an output of the Nature+Energy Project, funded by Research Ireland (12/RC/2302\_P2), industry partners (Wind Energy Ireland, NTR Foundation, ESB, SSE Renewables, Asper Investment Management, Ecopower, Energia, EnergyPro, Greencoat Renwables, Ørsted) and MaREI, the Research Ireland Research Centre for Energy, Climate and Marine Research and Innovation.

We thank our labellers Harry Hussey, David Kelly, Seán Ronayne and Mark Shorten, who annotated the majority of the audio files in the test set using NEAL. We also thank the on-site personnel and surveyors for carrying out recorder maintenance, i.e., data and battery transfers, as the wind farms are widely spread across Ireland.

Andrew Parnell’s work was supported by: the UCD-Met Éireann Research Professorship Programme (28-UCDNWPAI); a Research Ireland, Northern Ireland’s Department of Agriculture, Environment and Rural Affairs (DAERA), UK Research and Innovation (UKRI) via the International Science Partnerships Fund (ISPF) under Grant number [22/CC/11103] at the Co-Centre for Climate + Biodiversity + Water. For the purpose of Open Access, the authors have applied a CC BY public copyright licence to any Author Accepted Manuscript version arising from this submission.

\subsection*{Data and Code Availability}\label{ch:availability}

Code and documentation for training both generative models, all classification scripts are available at \url{https://github.com/gibbona1/SpectrogramGenAI}. The post-processing scripts, to generate the plots used in this paper, are also available here.  The large datasets of spectrograms, embeddings and labels are available at \url{https://zenodo.org/records/15729847}. 

\printbibliography

\appendix

\section*{Supplemental Information}

\begin{table}[!ht]
\centering
\begin{tabular}{|l|l|c|l|}
\hline
\textbf{Layer Type} & \textbf{Output Shape} & \textbf{Param \#} & \textbf{Description} \\
\hline
Input & [-1, 1, 256, 256] & - &Spectrogram input\\
Conv2d & [-1, 16, 256, 256] & 160 & 3x3 convolution, 16 filters \\
MaxPool2d & [-1, 16, 128, 128] & -- & 2x2 max pooling \\
Conv2d & [-1, 32, 128, 128] & 4,640 & 3x3 convolution, 32 filters \\
MaxPool2d & [-1, 32, 64, 64] & -- & 2x2 max pooling \\
Conv2d & [-1, 64, 64, 64] & 18,496 & 3x3 convolution, 64 filters \\
MaxPool2d & [-1, 64, 32, 32] & -- & 2x2 max pooling \\
Conv2d & [-1, 128, 32, 32] & 73,856 & 3x3 convolution, 128 filters \\
MaxPool2d & [-1, 128, 16, 16] & -- & 2x2 max pooling \\
Flatten & [-1, 32768] & -- & Flatten to 1D vector \\
Dropout & [-1, 32768] & -- & Dropout layer \\
Linear & [-1, 256] & 8,388,864 & Fully connected, 256 units \\
Dropout & [-1, 256] & -- & Dropout layer \\
Linear & [-1, 27] & 6,939 & Fully connected, 27 units \\
\hline
\textbf{Total} & & \textbf{8,492,955} & \\
\hline
\end{tabular}
\caption{Architecture of the Custom Model}
\label{tab:custom_architecture}
\end{table}

\begin{table}[!ht]
\centering
\begin{tabular}{|l|c|c|c|}
\hline
\textbf{Model} & \textbf{Total Params} & \textbf{Multiply-Adds (billions)} & \textbf{Est. Size (MB)} \\
\hline
ResNet18 & 11.19M & 2.39 & 92.94 \\
VGG16 & 134.37M & 20.30 & 648.40 \\
MobileNetV2 & 2.26M & 0.20 & 30.02 \\
Custom & 8.49M & 0.24 & 47.65 \\
Ensemble & 156.32M & 20.88 & 762.61 \\
\hline
\end{tabular}
\caption{Key Metrics for Each Model}
\label{tab:model_metrics}
\end{table}

\begin{figure}
     \centering
     \begin{subfigure}{\textwidth}   \includegraphics[width=\textwidth]{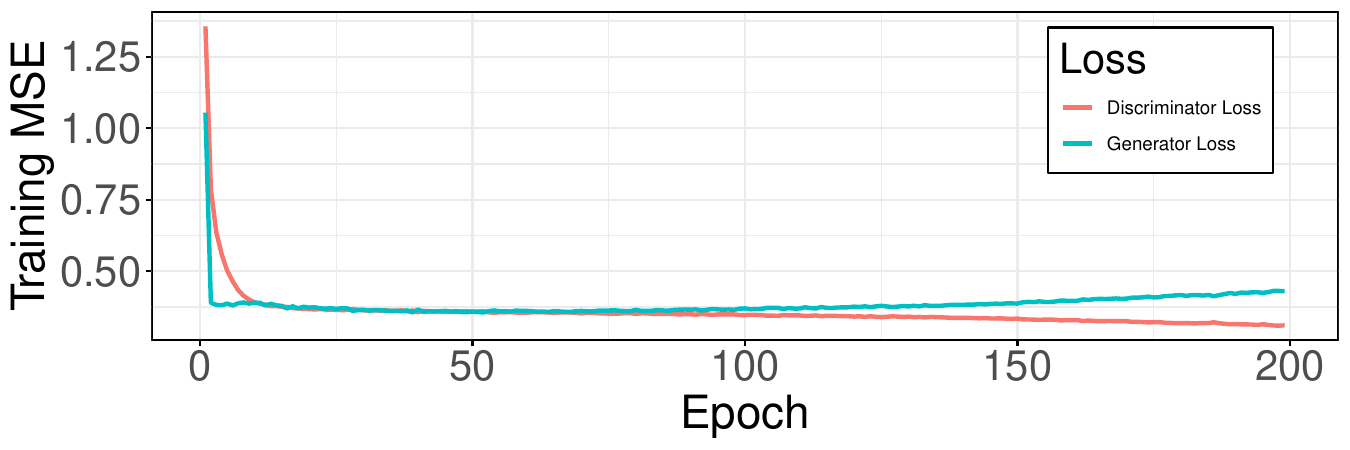}
     \caption{ACGAN Training Losses}
     \label{fig:gan_loss}
     \end{subfigure}
     \begin{subfigure}{\textwidth}
     \includegraphics[width=\textwidth]{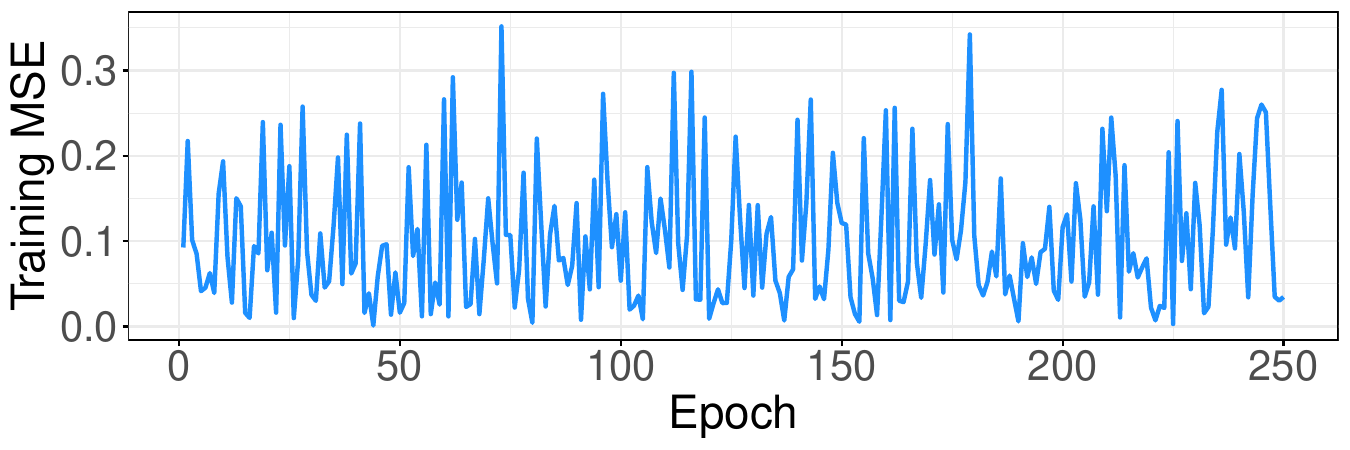}
     \caption{DDPM Training Loss}
     \label{fig:ddpm_loss}
     \end{subfigure}
    \caption{\textbf{Generative model training losses.} ACGAN has generator and discriminator. DDPM has MSE.
    }
    \label{fig:losses}
\end{figure}

\end{document}